\documentclass[prl,twocolumn,showpacs,preprintnumbers,amsmath,amssymb]{revtex4}

\usepackage{graphicx}
\usepackage{dcolumn}
\usepackage{bm}

\begin{document}

\title{A mass formula for baryon resonances}

\author{Eberhard Klempt}
 \affiliation{Helmholtz-Institut f\"ur Strahlen- und Kernphysik der
Universit\"at Bonn, Germany}

\date{\today}

\begin{abstract}
Light-baryon resonances with u,d, and s quarks only can be classified 
using the non-relativistic quark model. When we assign to
baryon resonances with total angular momenta {\rm J} intrinsic orbital 
angular momenta 
{\rm L} and spin {\rm S} we make the following observations: 
plotting the squared masses of the light-baryon resonances 
against these intrinsic orbital angular momenta {\rm L}, 
$\Delta^*$'s with even and odd parity can be described by the same
Regge trajectory. For a given {\rm L}, nucleon resonances with spin 
{\rm S}=3/2 are approximately degenerate in mass with $\Delta$ 
resonances of same total orbital momentum {\rm L}. To which total
angular momentum {\rm L} and {\rm S} couple has no significant 
impact on the baryon mass. 
Nucleons with spin 1/2 are shifted in mass; the
shift is - in units of squared masses - proportional to the component
in the wave function which is antisymmetric in spin and
flavor. Sequential resonances in the same partial wave are separated 
in mass square by the same spacing as observed in orbital angular
momentum excitations. Based
on these observations, a new baryon mass formula is proposed which
reproduces nearly all known baryon masses.
\end{abstract}


\maketitle
\newcommand{\bc}        {\begin{center}}
\newcommand{\ec}        {\end{center}}

Phenomenological analyses of transition energies between energy levels
of bound systems can provide deep insight into the 
underlying dynamics. The Balmer formula demonstrated 
that the interpretation of the hydrogen atom must be simple; the 
formula was given long before Bohr derived the famous model
which bares his name. Our 
understanding of nucleon-nucleon interactions was boosted by the 
discovery that the magic numbers in nuclear physics can be understood
in terms of a nuclear shell structure in the presence of strong
spin-orbit forces. And the analogy of the charmonium states with those
of positronium atoms provided not only evidence for the existence of a
new flavor but was also the final proof for the reality of quarks. 
In this letter we propose a new mass formula for light baryon resonances
which reproduces 81 of the 82 masses of baryons with known spin and parity
given by the Particle Data Group~\cite{Groom:2000in}.  
We assume that the baryon mass spectrum is due to the dynamics of
three constituent quarks and that a confinement interaction gives rise to
linear Regge trajectories \cite{Tang:2000tb}.
The study aims to identify
the dominant residual interactions between the constituent quarks.
The mass formula reads~\cite{Klempt}: 

\vspace*{-4mm}
\begin{equation}
\label{mass}
\rm  M^2 = M_{\Delta}^2 + \frac{n_s}{3}\cdot M_s^2  + 
a\,(L + N) - s_i \cdot I_{sym}
\end{equation}
where 
\vspace*{-6mm}
\bc
$\rm M_s^2 = \left(M_{\Omega}^2 - M_{\Delta}^2\right) $\\
$\rm s_i = \left(M_{\Delta}^2 - M_{\rm N}^2\right)$.
\ec
\noindent
$n_s$ is number of strange quarks in the baryon.
Mostly, baryon masses are assumed to increase linearly with the
number of strange quarks. We use in (\ref{mass}) a quadratic
dependence for sake of simplicity. The model has no parameter to
account for the $\Lambda -\Sigma$ mass difference~\cite{Loring:2001ky}.
L is the total intrinsic orbital angular momentum, which we have to
assign to each baryon resonance. N is the radial
excitation quantum number; L+2N
gives the harmonic-oscillator band. 
$\rm M_{N}, M_{\Delta}, M_{\Omega}$ are input parameters taken from
PDG.
$a=1.142$/GeV$^2$ is the Regge slope determined from the series of 
light (isoscalar and isovector) mesons with quantum numbers 
J$^{\rm PC} = 1^{--}, 2^{++}$, $3^{--}$, $4^{++}$, $5^{--}$, 
$6^{++}$.
$\rm I_{sym}$ is the fraction of the 
wave function (normalized to the nucleon wave function) 
antisymmetric in spin and flavor. It depends on the SU(6) flavor wave function

\vspace*{-4mm}
\bc
\renewcommand{\arraystretch}{1.1}
\begin{tabular}{lrcc}
$\rm I_{sym}  = $&1.0& for S=1/2 and & for octet baryons in 56-plets; \\
$\rm I_{sym}  = $&0.5& for S=1/2 and & for octet baryons in 70-plets;\\
$\rm I_{sym}  = $&1.5& for S=1/2 and & for singlet baryons; \\
$\rm I_{sym}  = $&  0& otherwise.
\end{tabular}
\renewcommand{\arraystretch}{1.0}
\ec
\vspace*{-2mm}
\par 
For a quantitative comparison of our mass formula (\ref{mass}) 
with the experimental masses of the light-baryon resonances, 
central values and their uncertainties need to be defined. 
As mass value of a resonance 
we take - when given - the central value of the interval suggested 
by the Particle Data Group.
We do not take experimental uncertainties of the mass determination
into account, since they are only given for well established resonances.
Instead, we use a simple estimate 
based on contributions from the hadronic width and a model error.
It is well known that hadronic effects like opening thresholds,
virtual decays and mixing with other states 
may result in mass shifts. To account for these effects we allow
for an error of one quarter of the 
hadronic width of a resonance. A constant model error of 30 MeV
is added quadratically to give the total error $\sigma_{\rm M}$.
Since the measured widths show a wide spread and are often 
rather inaccurate, we use $\Gamma = Q/4$ as width estimate 
where $Q$ is the largest kinetic energy accessible in hadronic 
decays of the resonance. Our estimated uncertainties 
vary (for N and $\Delta$ 
resonances) from 40 MeV at 1500 MeV to 120 MeV at 3 GeV. 
Note that experimental uncertainties 
in the mass determination are often in the same range. 
\par
According to eq. (\ref{mass}),
the squared baryon masses depend linearly on the intrinsic orbital
angular momentum {\rm L}. 
Measured is of course only the total angular momentum {\rm J}.
We identify multiplets with intrinsic spin 3/2 using the following
criteria: first, we identify 'stretched' states with
J=L+S; L=0,1,..,6 and S=3/2, i.e.~resonances with quantum 
numbers $\rm J^P = 3/2^+, 5/2^-, 7/2^+, 9/2^-$, $11/2^+, 
13/2^-, 15/2^+$. These are shown in Table~\ref{positive} 
in the last column. Omitted are the decuplet ground states (L=0) 
which also fall into this category. 
\par
In our eq. (\ref{mass}) we do not account for spin-orbit forces, assuming
they are small or vanishing. Therefore, we collect all resonances 
of a spin 3/2 multiplet from a mass 
window  (here we chose $\pm\sqrt{2}\sigma_{\rm M}$) 
around the same stretched state requiring the same parity.
In the non-relativistic quark model we expect single resonances for
L=0 (the ground states), triplets for L=1 and quartets for higher L.
The multiplet structure is clearly visible in Table~\ref{positive},
even though the multiplets are not complete, supporting our assumption
in eq. (\ref{mass}) of small or vanishing spin-orbit forces.  
\par
\begin{table}[h!]
\caption{Baryon resonances assigned to S=3/2 multiplets. Baryon masses
depend only weakly on the orientation of the spin relative to the
orbital angular momentum:  spin-orbit
forces are small ($\vec{\rm L}\cdot\vec{\rm S}\sim 0$). 
Missing states are marked by a - sign.}
{
\renewcommand{\arraystretch}{1.3}
\hspace{-10mm}
\bc
\begin{tabular}{ccccc}
\hline
\hline
L & J=L-3/2 & J=L-1/2 & J=L+1/2 & J=L+3/2\\
1 &
& N$_{1/2^-}(1650)$
& N$_{3/2^-}(1700)$
& N$_{5/2^-}(1675)$ \\
1 &
& $\Delta_{1/2^-}(1900)$
& $\Delta_{3/2^-}(1940)$
& $\Delta_{5/2^-}(1930)$ \\
1 & 
& $\Lambda_{1/2^-}(1800)$
&\ - \
& $\Lambda_{5/2^-}(1830)$ \\
1 & 
& $\Sigma_{1/2^-}(1750)$
&\ - \
& $\Sigma_{5/2^-}(1775)$\\
2 & \ - \
& N$_{3/2^+}(1900)$
& N$_{5/2^+}(2000)$
& N$_{7/2^+}(1990)$ \\
2 &  $\Delta_{1/2^+}(1910)$
& $\Delta_{3/2^+}(1920)$
& $\Delta_{5/2^+}(1905)$
& $\Delta_{7/2^+}(1950)$ \\
2 & \ - \
&\ - \ 
& $\Lambda_{5/2^+}(2110)$
& $\Lambda_{7/2^+}(2020)$ \\
2 & \ - \
& $\Sigma_{3/2^+}(2080)$
& $\Sigma_{5/2^+}(2070)$
& $\Sigma_{7/2^+}(2030)$ \\
3 & \ - \
& N$_{5/2^-}(2200)$
& N$_{7/2^-}(2190)$
& N$_{9/2^-}(2250)$\\
3 & \ - \
& $\Delta_{5/2^-}(2350)$ 
&\ - \
& $\Delta_{9/2^-}(2400)$\\
4 & \ - \ 
& $\Delta_{7/2^+}(2390)$
& $\Delta_{9/2^+}(2300)$
& $\Delta_{11/2^+}(2420)$\\
5 & \ - \
&\ - \
&\ - \
&$\Delta_{13/2^-}(2750)$\\
6 & \ - \
&\ - \
&\ - \
&$\Delta_{15/2^+}(2950)$\\
\hline
\hline
\end{tabular}
\ec
\renewcommand{\arraystretch}{1.0}
}
\label{positive}
\end{table}
\par
Also quantitatively the comparison of 
our mass formula (\ref{mass}) with the light-baryon resonances 
with spin assignment S=3/2 (see Table~\ref{positive}) is doing well:
We get a $\chi^2=23.6$ for 31 data points.
\par
\begin{figure*}
\includegraphics[width=0.8\textwidth]{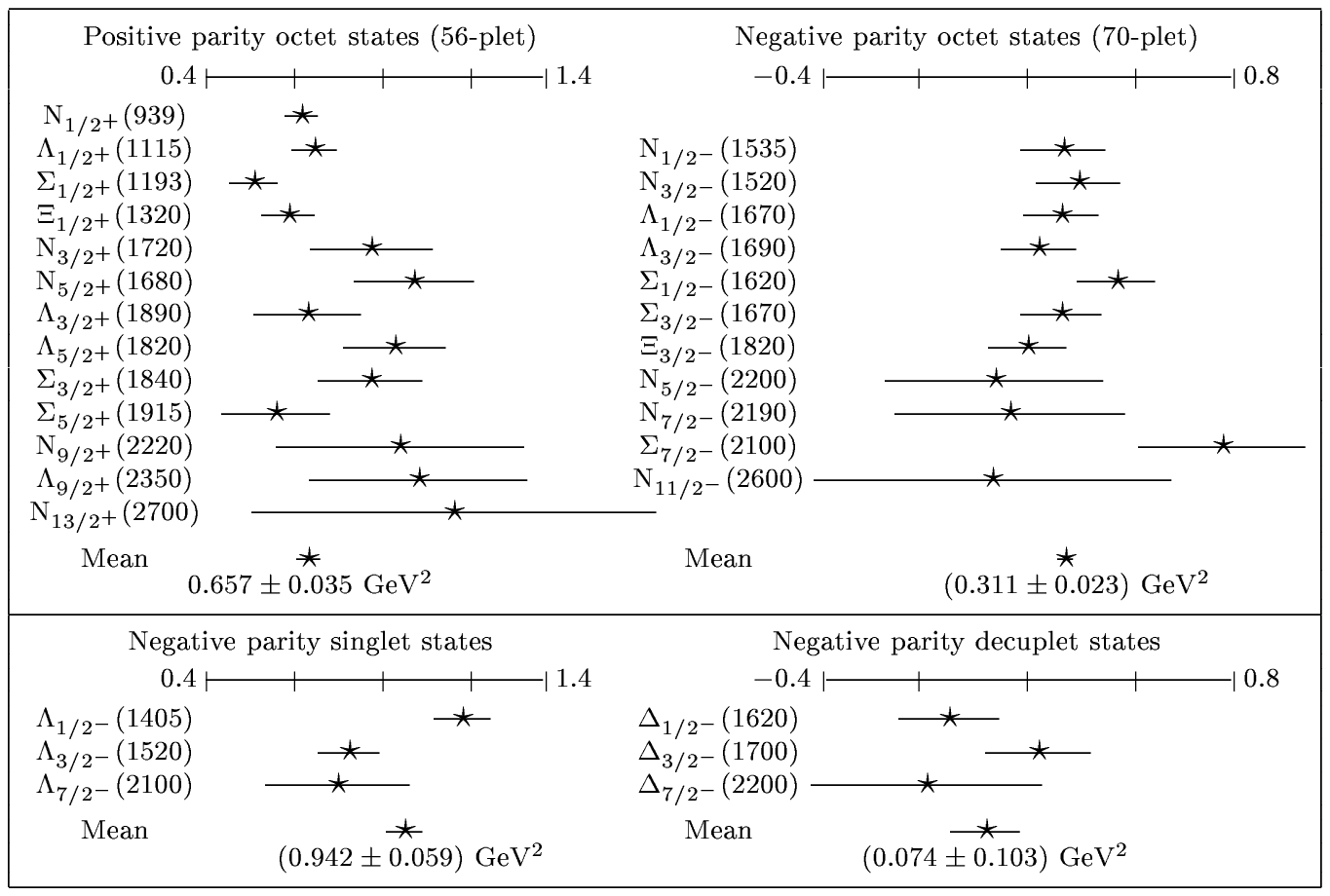}
\caption{\label{instant}Mass square shift (in GeV$^2$) 
of spin-1/2 baryons w.r.t. the Regge trajectory 
$\rm  M^2 = M_{\Delta}^2 + {n_s}/3\cdot M_s^2  + a\,(L + N)$
defined by
baryons with S=3/2 (hyperfine splitting).  
The mass shifts scale as 1\,:\,1/2\,:\,3/2\,:\,0 times
$\rm M_{\Delta}^2 - M_{\rm N}^2$ as we proposed in
mass formula (\ref{mass}).} 
\end{figure*}

We now turn to a discussion of spin 1/2 resonances. 
The lowest-mass spin-1/2 states have intrinsic L=0, 
positive parity and belong to an octet in the 56-plet 
representation. We now search for doublets of nearly 
mass-degenerate states with J=L$\pm$1/2. Doublets are 
observed for L=1,2, and 3; for larger L only one state 
with L+1/2 is known. The spin 1/2 states are collected in 
Fig. \ref{instant}, grouped according to 
their SU(6) classification. 
The positive parity octet states have a shift
in squared mass relative to the Regge trajectory of $0.657\pm 0.035$
GeV$^2$. This value is compatible with the $\Delta_{3/2^+}(1232)$-N 
mass square difference ($0.636$ GeV$^2$). The negative-parity
octet resonances undergo a mass shift of $(0.311\pm 0.023)$ GeV$^2$,
consistent with 1/2 of the $\Delta_{3/2^+}(1232)$-N 
mass square difference. We have also included the N$_{5/2^-}(2200)$
and N$_{7/2^-}(2190)$ from Table~\ref{positive} here, since their
intrinsic spin assignment (S=3/2 or 1/2) is ambiguous.

\par
The $\Lambda_{1/2^-}(1405)$ and
$\Lambda_{3/2^-}(1520)$ with their low masses are assigned to the
SU(6) singlet system; the two states $\Lambda_{1/2^-}(1670)$
and $\Lambda_{3/2^-}(1690)$ form then the spin doublet of the 70-plet
octet, and the $\Lambda_{1/2^-}(1800)$ and $\Lambda_{5/2^-}(1830)$
an incomplete spin-triplet (also belonging to a 70-plet). 
The $\Lambda_{7/2^-}(2100)$ is the lowest $\Lambda$ resonance with L=3;
we assign it to the SU(6) singlet system because of its mass.
The assignment is thus {\it ad hoc} as long 
as its octet partner (predicted by (\ref{mass}) at a mass of 2318 MeV) 
has not been found. 
These three singlet resonances have a large mass shift down from the Regge
trajectory of $(0.942\pm 0.059)$ GeV$^2$ or 3/2 
times the $\Delta_{3/2^+}(1232)$-N 
mass difference. 
\par
There is one doublet of negative-parity $\Delta$ states, the 
$\Delta_{1/2^-}(1620)$ and $\Delta_{7/2^-}(1700)$. 
In addition we assign
the $\Delta_{7/2^-}(2200)$ to the lowest-mass state  with L=3
and S=1/2. 
It could also form a spin-3/2 quartet with the two other
resonances $\Delta_{5/2^-}(2350)$ and $\Delta_{9/2^-}(2400)$. 
However, the  $\Delta_{7/2^-}(2200)$ does not fall into the
$\pm\sqrt{2}\sigma_{\rm M}$ 
corridor, hence we do not accept this as spin 3/2 state.
The mean mass shift of the three remaining negative parity
decuplet $\Delta$ states
relative to the Regge trajectory is
$(0.074\pm0.103)$  GeV$^2$ and compatible with zero.
\par
Summarizing the S=1/2 states, we observe a reasonable agreement 
with the experimental masses with eq. (\ref{mass}) resulting
in a $\chi^2$ contribution of 43.3 for 29 d.o.f.
Especially the description of the 
deviation of (\ref{mass}) from the Regge trajectory
by the additional symmetry term (last term in (\ref{mass}))
is nicely confirmed. 

In Eq.~(\ref{mass}), radial excitations 
are supposed to have the same mass spacing (per unit of excitation
number) as orbital angular momentum excitations. In Table~\ref{radial} 
we list resonances belonging to one partial wave, 
and their mass square differences. The differences are of the
order of 1.1 GeV$^2$, not incompatible with the spacing per unit
of L. The 14 new data points contribute 
$\delta\chi^2$=17.8.
This observation is the basis of the L+N dependence in
(\ref{mass}). Table~\ref{radial} may contain some positive-parity 
resonances with L=2, S=3/2 with ambigous assignments. 
\par
\begin{table}
\caption{Excitations of baryon resonances having the same quantum
numbers. The mean value per excitation is $(1.081\pm 0.036)$ GeV$^2$, 
to be compared to the 1.142 GeV$^2$ from the fit
to the meson Regge trajectory.}
\bc
\renewcommand{\arraystretch}{1.2}
\begin{tabular}{cccc}
\hline
\hline
Baryon   &$\delta\rm M^2$(GeV$^2$)&Baryon   & $\delta\rm M^2$(GeV$^2$) \\  
N$_{1/2^+}(939)$  &               &$\Delta_{3/2^+}(1232)$ &                 \\
N$_{1/2^+}(1440)$      & $1 (1.18\pm 0.11)$ 
&$\Delta_{3/2^+}(1600)$& $1 (1.04\pm 0.15)$   \\
N$_{1/2^+}(1710)$      & $2 (1.02\pm 0.18)$ 
&$\Delta_{3/2^+}(1920)$& $2 (1.08\pm 0.24)$   \\
N$_{1/2^+}(2100)$      & $3 (1.18\pm 0.29)$ 
&                      &                 \\
\hline
$\Lambda_{1/2^+}(1115)$&             
&$\Sigma_{1/2^+}(1193)$&               \\  
$\Lambda_{1/2^+}(1600)$& $1 (1.24\pm 0.10)$
&$\Sigma_{1/2^+?}(1560)$&$1 (1.04\pm 0.10)$  \\ 
$\Lambda_{1/2^+}(1810)$& $2 (0.98\pm 0.15)$
&$\Sigma_{1/2^+}(1880)$& $2 (1.06\pm 0.11)$ \\ 
\hline
N$_{1/2^-}(1535)$      &             
&N$_{3/2^-}(1520)$     &               \\
N$_{1/2^-}(2090)$      &$2 (1.01\pm 0.31)$
&N$_{3/2^-}(2080)$     &$2 (1.01\pm 0.31)$  \\
$\Delta_{1/2^-}(1620)$ &             
&$\Delta_{3/2^-}(1700)$&               \\
$\Delta_{1/2^-}(1900)$ &$1 (0.99\pm 0.24)$
&$\Delta_{3/2^-}(1940)$&$1 (0.87\pm 0.24)$  \\
$\Delta_{1/2^-}(2150)$ &$2 (1.00\pm 0.34)$
&                      & \\
\hline
                       &                &
$\Lambda_{3/2^-}(1670)$& \\  
                       &                &
$\Lambda_{3/2^-}(2325)$&$2 (1.31\pm 0.27)$\\ 
$\Sigma_{1/2^-}(1620)$ &               &
$\Sigma_{3/2^-}(1670)$ & \\  
$\Sigma_{1/2^-}(2000)$ &$1 (1.37\pm 0.18)$  & 
$\Sigma_{3/2^-}(1940)$ &$1 (0.97\pm 0.17)$ \\ 
\hline
\hline
\end{tabular}
\renewcommand{\arraystretch}{1.0}
\ec
\label{radial}
\end{table}
\par
So far, we have included all baryon resonances of known spin-parity
except a few special cases.  The $\Sigma_{3/2^-}(1580)$ has two stars
in the PDG notation, but it is very low in mass and does possibly not
exist~\cite{Manley:2002ue}. We disregard this resonance.
The $\Delta_{5/2^+}(2000)$ has two mass entries, at 1752 MeV
and 2200 MeV, respectively. Using the higher mass value, it can be
identified as radial excitation of the $\Delta_{5/2^+}(1905)$ but this
is clearly a speculation. There remain three states to be discussed,
the $\Delta_{1/2^+}(1750)$ with one *, the 
$\Sigma_{1/2^+}(1660)$ (***) and the $\Sigma_{1/2^+}(1770)$ (*).
Radial excitations of the $\Sigma_{3/2^+}(1385)$ are not necessarily
in a 56-plet (then they have $3/2^+$); they can also fall into a 70-plet. 
In this case they have spin 1/2. The difference in squared mass
between the $\Sigma_{1/2^+}(1770)$ and the $\Sigma_{3/2^+}(1385)$
is 1.21 GeV$^2$, compatible with the other values in
Table~\ref{radial}. 
The $\Delta_{1/2^+}(1750)$ could be an analogous state;
in this case the mass square difference is uncomfortably large,
1.54 GeV$^2$; however the $\Delta_{1/2^+}(1750)$ is a one * resonance
only. Likewise, the $\Sigma_{1/2^+}(1660)$ could be an octet radial
excitation belonging to the SU(6) 70-plet. The mass difference to the
first radial excitation in the 56-plet, possibly the $\Sigma(1560)$, is
0.322 GeV$^2$, nearly identical to the other splittings between
resonances belonging to the 56 or 70-plet. So, while the resonances
discussed in this last paragraph cannot be used to validate the mass
formula (\ref{mass}), they are nevertheless consistent 
with it when appropriate quantum numbers are assigned. These four
states and the two remaining decuplet ground states (the $\Delta$ and
$\Omega$ masses are used as input parameters) contribute 
$\delta\chi^2$=7.1. 
\par
In summary we compared 81 resonances to their masses according to the
values summarized by the Particle Data Group and obtain a
$\chi^2 = 91.7$ for 78 degrees of freedom.
\par
We now discuss consequences for our understanding of the baryon
mass spectrum. 
The mass formula (\ref{mass}) contains the orbital angular momentum as
decisive quantity for baryon masses. The orbital angular momentum
is the sum $\vec{\rm L}=\vec l_{\rho}+\vec l_{\lambda}$
of two orbital angular momenta associated with the two
generalized coordinates of the three-particle system. 
All resonances are compatible with  either ${\rm L}=l_{\rho}$ or
${\rm L}=l_{\lambda}$. A dynamical reason for this selection rule
is not known; the question is related to the {\it missing resonance
problem}. 
\par
Baryon resonances are classified according to the non-relativistic quark
model. Doublets and quartets are clearly identified in the mass
spectrum. The mass formula (\ref{mass}) does not include 
spin-orbit interactions. 
The proton spin puzzle underlines that our 
understanding of the dynamical role of the quark spin in baryons 
is not sufficient to exclude the possibility that 
spin-orbit interactions play no or little role in the baryon mass
spectrum. 
\par
The second point resulting from this analysis is the energy gap of
radial excitations. In the harmonic oscillator approximation,
the first radial excitations are found in the second excitation band;
the anharmonicity due to the confinement potential - supposed to be
linear - shifts its mass down but not low enough to hit the mass of
the Roper resonance at 1440 MeV or the $\Delta_{3/2^+}(1600)$. 
Table~\ref{radial} shows a large number of recurrencies (17)
which all give a small mass shift per increase in radial excitation
number. Bijker {\it et al.}~\cite{Bijker:1994yr} have used an 
algebraic approach
to describe baryon resonances. For them, the lowest recurrencies are
one-phonon excitations and not two-phonon excitations as in the
harmonic oscillator model. 
\par
Spin-spin interactions depend on the SU(6)
symmetry of the baryon wave function.
The symmetry term in (\ref{mass}) acts 
only for octet and singlet baryons (which have a component antisymmetric
w.r.t. the exchange of two quarks) with spin 1/2 (which also has 
a component antisymmetric w.r.t. the exchange of two quarks). 
This latter component is reduced by a factor 2 in wave
functions belonging to SU(3) octets within the SU(6) 70-plet. Of
course, the overall wave functions in 56-plets and 70-plets have the
same symmetry. Loosely speaking, in baryons with odd angular momentum, 
part of the antisymmetry is found in the spacial wave function. 
The $\Lambda$ resonances in the SU(6) singlet have negative parity,
too. But now, all three quark pairs are antisymmetric in flavor w.r.t.
exchange of two quarks. This gives the factor 3/2 enhancement of the
symmetry contribution. 
Decuplet baryons or baryons with spin 3/2 do not have a wave function
which is antisymmetric w.r.t. the exchange of two quarks both in spin
and in flavor. They all fall onto the main Regge trajectory. 
\par
We thus need an interaction which gives rise to a mass shift
proportional to the fraction of the wave function which is
antisymmetric w.r.t. the exchange of two quarks both in spin and in
flavor. This is a selection rule which holds for instanton-induced
interactions~\cite{Shuryak:1989bf}. The success of the eq.~(\ref{mass})
provides therefore strong support that instanton-induced interactions
play a decisive role for the spectrum of baryon resonances 
and are responsible for the hyperfine splitting. 
Interactions ascribed to one-gluon exchange  
can - at least to first order - be neglected.
\par
The most model-discriminating masses are those 
of the negative-parity $\Delta$ resonances above 1.8 GeV. 
Capstick~\cite{Capstick:1986bm}
finds them at about 2.1 GeV, L\"oring {\it et al.}~\cite{Loring:2001kx} at
2.2 GeV. Bijker {\it et al.}~\cite{Bijker:1994yr} fit the data (with 11 parameters) 
and find 1.9 GeV, in agreement with data. 
In Glozman {\it et al.}~\cite{Glozman:1998ag}
only the lower-mass states are calculated. The mass formula
(\ref{mass}) yields 1.95 GeV. The least established 
$\Delta_{3/2^-}$ resonance is predicted to dominate the
reaction $\gamma$p$\rightarrow\Delta_{3/2^-}\rightarrow
\Delta_{3/2^+}(1232)\eta$ where the latter decay is in
S-wave. Experiments along these lines are presently performed
at ELSA~\cite{Crede}.
\par
We have shown that the spectrum of baryon resonances can be
described successfully by a very simple mass formula. The squared
masses increase linearly with the intrinsic orbital angular momentum
between the constituent quarks, radial excitations have the same
spacings as orbital excitations. Instanton-induced interactions reduce the
masses whenever a component of the baryonic wave function is sensitive 
to their action. Gluon exchange leads to no significant contributions.


\begin{acknowledgments}
We wish to acknowledge discussions with 
D. Diakonov, K. Goeke, B. Metsch, H. Petry, B. Schoch
and Chr. Weinheimer.
\end{acknowledgments}
\bibliography{regge_prl}
\end{document}